# TRANSFORMATION OF AN ELECTROMAGNETIC FIELD INTO GRAVITATIONAL FIELD IN A MODEL OF EXTENDED SPACE: PREDICTION AND EXPERIMENT.

Tsipenyuk D.Yu.[*]

General Physics Institute of the Russian Academy of Sciences,
119991, Vavilova str.38, Moscow, Russia

## 1. Introduction

In the previous works [1-7] the generalization of a Einstein's special theory of relativity on 5-dimensional extended space with the metric (+ ; -, -, -, -).

Was constructed the model that unified electromagnetic and gravitational interactions. Models equations are formulated in extended space, in that space to usual space (x, y, z) and time t coordinates the fifth coordinate s was added.

This 5-th coordinate has geometrical meaning - interval in space of the Minkowski. Physically we connect it with an index of refraction n. It is supposed, that in case of the motion in trajectory with a variable n the rest-mass of particles varies. This cause a modification of a gravitational field, created by these particles.

In particular, particles with a zero mass (photons) gain a nonzero mass and begin to be for a source of a gravitational field, when move from blank space with n = 1 to the medium with n> 1. Offered in [4] unified set of equations is called to describe such processes.

In this space the mechanics a mass point [1,2,6] and electrodynamics [2,3] is constructed. Lienar-Vihert potentials appropriate to such model also are considered [5]. The properties of solutions, adequate them, of an extended set of equations of the Maxwell are analyzed [5].

The gravitational effects in extended space were also considered: the second space velocity, red shift and deflection of starlight, retardation of radar-echo from Mars [7]. It was shown that it is possible to receive the same formula (which are received in a general theory of relativity for calculation of magnitude of these effects) completely in another way within the framework of a model of extended space. For this purpose it was supposed, that the gravitational field create in space some index of refraction n. Refraction index n depends on strength of this field. This index of refraction n determines motion both photons and massive particles in this space. Then the technique of turns in extended space [1-4] was applied.

Basic outcomes obtained in the previous works are briefly reduced below (also some results in electronic versions could be found in [11,13]).

It was offered such generalization of special theory of relativity, which takes into account processes under which a mass of a particle m also would be a variable. As a mass of a particle m we, following the recommendations of the review [8], shall understand it a rest-mass, which is a Lorenz scalar. For this purpose first of all we shall construct the extension (1 + 3) - dimensional space of the Minkowski $M(T;\vec{X})$ on (1 + 4) - dimensional space $G(T;\vec{X},S)$. This space we will name as extended space.
As the 5-th additional coordinate we will use the interval S that already exists in Minkowski space

$$s^2 = (ct)^2 - x^2 - y^2 - z^2. \qquad (1)$$

---

[*] E-mail: tsip@kapella.gpi.ru

The interval S is saved for usual Lorentz transformations in the Minkowski space $M(T;\vec{X})$ but varies under turns in extended space $G(T;\vec{X},S)$. Thus, the Minkowski space $M(T;\vec{X})$ -is a cone in extended space $G(T;\vec{X},S)$. In this space saved only

$$s^2 - (ct)^2 - x^2 - y^2 - z^2 = \text{const.} \qquad (2)$$

It is known, that an energy, momentum and mass of a free particle are connected by a relation [9].

$$E^2 - c^2 p_X^2 - c^2 p_Y^2 - c^2 p_Z^2 - m^2 c^4 = 0. \qquad (3)$$

This is analog of a relation (1) in space $G'(E;\vec{P},M)$, which conjugate to space $G(T;\vec{X},S)$. The mass m is conjugate to the interval s. The transformations of the Lorentz change an energy of a particle E and it impulse $\vec{p}$, but leave by a constant a mass m. Transformations in extended space complementary to transformations of the Lorentz, change also mass of a particle, leaving a constant only form (4).

$$E^2 - c^2 p_X^2 - c^2 p_Y^2 - c^2 p_Z^2 - m^2 c^4 = \text{const.} \quad (4)$$

In Minkowski space $M(E;\vec{P})$ for free particles corresponds a 4-component vector of energy - momentum $\vec{p} = (\frac{E}{c}, p_x, p_y, p_z)$, which components are connected by a relation (3). In dependence from that, does the mass m of a particle is equal to zero or not, the point appropriate to this vector, lies either on a cone, or on a hyperboloid in space $M(E;\vec{P})$.

In extended space $G'(E;\vec{P},M)$ 5- component vector of energy - momentum- mass $\bar{p} = (\frac{E}{c}, p_x, p_y, p_z, mc)$ correspond to free particles. Components of 5-vector connect by a relation (3). Now mass m is not any more constant exterior parameter, and becomes variable. All these 5- vectors are isotropic and the points, appropriate to them, lie on a cone (3) in space $G'(E;\vec{P},M)$. The particles which are in an exterior field, are described by nonisotropic vectors lying on hyperboloids (2) [1,2].

In Minkowski space all particles are separated on two types. At first massive, described by a mass m. Secondly massless (photons) described by frequency $\omega$. In our model to a massive particle at rest corresponds the 5-vector of energy - momentum-mass: $\bar{p}_m = (mc;\vec{0}, mc)$.

For photon that moves in blank space with a velocity c in direction $\vec{k}$ correspond the 5-vector of energy - momentum-mass

$$\bar{p}_{\hbar\omega} = (\frac{\hbar\omega}{c}; \frac{\hbar\omega}{c}\vec{k}, 0).$$

It is easy to see, that both of above 5-vectors are isotropic. Thus in our model for free point particles the isotropic vector corresponds in extended space $G'(E;\vec{P},M)$.

In space $G(T;\vec{X},S)$ there are new turns additionally to Lorentz transformations. These new turns intermix space (x, y, z) and time t coordinates with a new coordinate s. In a conjugate space $G^{'}(E;\vec{P},M)$ these turns translate energy and momentum in a mass and back [1,2].

There is one more 4-vector, with which works in special theory of relativity - 4-vector of a current. In a usual electrodynamics this 4-vector of a current, is received from a 4-vector of an energy - momentum of a charged particle by division it on a rest-mass of a particle m and multiplication on a denseness of a charge.

Such current is a source of an electromagnetic field, which is described by a 4-vector potential. Each component of a current can be considered as a source, which corresponds to one of the components of a vector potential.

In our model particle corresponds the 5-vector of energy - momentum-mass. The fifth component of the 5-vector expressed through a rest-mass of a particle. For this reason it is seems natural to connect it with a gravitational field. We want to have a current, which simultaneously would be a source both for electromagnetic and gravitational fields. Therefore, we multiply this 5-vector on a denseness of a charge. In a traditional statement the electromagnetic current is described by a 4-vector in space of the Minkowski. It is necessary to transform a 4-vector of a current to a 5-vector. As well as in case of a vector of energy - momentum we demand that this vector will be isotropic. Thus, we obtain [1,3].

$$\overline{\rho} = (\rho, \vec{j}, j_s) = (\frac{\rho_0 c}{\sqrt{1-\beta^2}}, \frac{\rho_0 \vec{v}}{\sqrt{1-\beta^2}}, \rho_0 c), \quad \beta^2 = \frac{v^2}{c^2}. \qquad (5)$$

Here $\rho_0$ (t, x, y, z) - denseness of an electrical charge in the rest system of coordinates in a point (t, x, y, z). The denseness of a charge is invariant concerning Lorentz transformations and is analog of a rest-mass of a particle.

Magnitude $\vec{v} = (v_X(t,x,y,z), v_Y(t,x,y,z), v_Z(t,x,y,z))$ is local velocity of a denseness of charges. At such 5-current first four components are in a model of extended space $G(T;\vec{X},S)$ are source of an electromagnetic field. Fifth component is a source of a gravitational field. Such separation has a place in that case, when there are no processes changing rest-masses of particles. If such processes go, these two fields are integrated in a uniform electromagnetic-gravitational field.

The additional transformations, which are available in, extended, space $G(T;\vec{X},S)$, can changes magnitude $\rho_0$ similarly to that as they change a rest-mass m. Thus, within the framework of our model variables are both mass and charge of a particle.

In extended space $G(T;\vec{X},S)$ the current (5) is a source of a field, which is described by a 5-vector potential

$$(\varphi, \vec{A}, A_4) = (A_t, A_X, A_Y, A_Z, A_S) \qquad (6)$$

Potential (6) and current are connected by the equations [1,3]

$$\Diamond A_t = -4\pi \cdot \rho \qquad (7)$$

$$\Diamond \vec{A} = -\frac{4\pi}{c} \cdot \vec{j} \qquad (8)$$

$$\Diamond A_s = -\frac{4\pi}{c} \cdot j_s \qquad (9)$$

Here
$$\Diamond = \frac{\partial^2}{\partial s^2} + \frac{\partial^2}{\partial x^2} + \frac{\partial^2}{\partial y^2} + \frac{\partial^2}{\partial z^2} - \frac{1}{c^2}\frac{\partial^2}{\partial t^2} \qquad (10)$$

The field appropriate to a potential (6) contains in addition of usual electrical and magnetic components also additional components. These components reflect that fact, that during interaction the charge of particles can vary.

On potentials $(A_t, A_X, A_Y, A_Z, A_S)$ it is possible to construct a stress tensor of this field

$$F_{ik} = \frac{\partial A_i}{\partial x_r} - \frac{\partial A_k}{\partial x_i}; i,k = 0,1,2,3,4.$$

$$\|F_{ik}\| = \begin{pmatrix} 0 & -E_X & -E_Y & -E_Z & -Q \\ E_X & 0 & -H_Z & H_Y & -G_X \\ E_Y & H_Z & 0 & H_X & -G_Y \\ E_Z & -H_Y & H_X & 0 & -G_Z \\ Q & G_X & G_Y & G_Z & 0 \end{pmatrix} \quad (11)$$

Here new fields $Q$ and $\vec{G}$ have come in a stress tensor

$$Q = F_{40} = \frac{\partial A_4}{\partial x_0} - \frac{\partial A_0}{\partial x_4} = \frac{\partial A_S}{c \partial t} - \frac{\partial \varphi}{\partial s}. \quad (12)$$

$$G_X = F_{41} = \frac{\partial A_4}{\partial x_1} - \frac{\partial A_1}{\partial x_4} = \frac{\partial A_S}{\partial x} - \frac{\partial A_X}{\partial s}. \quad (13)$$

$$G_Y = F_{42} = \frac{\partial A_4}{\partial x_2} - \frac{\partial A_2}{\partial x_4} = \frac{\partial A_S}{\partial y} - \frac{\partial A_Y}{\partial s}.$$

$$G_Z = F_{43} = \frac{\partial A_4}{\partial x_3} - \frac{\partial A_3}{\partial x_4} = \frac{\partial A_S}{\partial z} - \frac{\partial A_Z}{\partial s}.$$

The expression for the force that affect on a particle (which has a charge e and mass m) in a field (11) was also found. Let's think, that in a system K', moved in extended space $G(T;\vec{X},S)$ together with a charge, charge is affected by the force which is given by a 4-vector $F' = (e\vec{E}', eQ')$. In such reference system equations of motion looks like

$$\frac{d\vec{p}}{dt} = e\vec{E}'; \quad \frac{dp_S}{dt} = eQ'. \quad (14)$$

In the case of transition in other system of reference with the help of turns in space $G(T;\vec{X},S)$ fields (11) will be transformed on rules established in [3].

Let's reduce the formulas noted with use of parameters $(\vec{v}, v_S, \vec{u})$. These parameters sets transition from one reference system to other [3,5].

1) $(T\vec{X})$ - this turns are characterized by a velocity $\vec{v}$. In this case, fields (11) will be transformed as follows

$$\vec{E}' = \vec{E} + \frac{1}{c}[v,\vec{H}], \qquad \vec{G}' = \vec{G} - \frac{1}{c}\vec{v}Q, \quad (15)$$

$$\vec{H}' = \vec{H} - \frac{1}{c}[v,\vec{E}], \qquad Q' = Q - \frac{1}{c}(\vec{v},\vec{G}).$$

2) (TS) – this turns are characterized by a velocity $v_s$ along a coordinate S. In this case fields (11) will be transformed as follows

$$\vec{E}' = \vec{E} + \frac{v_s}{c}\vec{G}, \qquad \vec{G}' = \vec{G} + \frac{v_s}{c}\vec{E}, \qquad (16)$$
$$\vec{H}' = \vec{H}, \qquad Q' = Q.$$

3) (S$\vec{X}$) - this turns are characterized by a parameter $\vec{u}$. It is a vectorial parameter that describes a modification of the index of refraction n during motion in direction of a vector $\vec{u}$. In this case fields (11) will be transformed as follows

$$\vec{E}' = \vec{E} - \vec{u}Q, \qquad \vec{G}' = \vec{G} + [\vec{u},\vec{H}], \qquad (17)$$
$$\vec{H}' = \vec{H} + [\vec{u},\vec{G}], \qquad Q' = Q + \frac{1}{c}(\vec{u},\vec{E}).$$

Let's apply these transformations to a system (14) and find a Lorentz force effecting on a moving particle. If we choose the sequence of transformations $(TS) + (T\vec{X}) + (S\vec{X})$, then the equations (14) change to

$$\frac{d\vec{p}}{dt} = e(\vec{E} - \frac{v_s\vec{v}}{c^2}(\vec{u},\vec{E}) - (\vec{u} + \frac{v_s\vec{v}}{c^2})Q +$$
$$+ \frac{1}{c}[\vec{v} + v_s\vec{u},\vec{H}] + \frac{1}{c}(v_s - (\vec{u},\vec{v}))\vec{G} + \frac{1}{c}\vec{u}(\vec{v},\vec{G})). \qquad (18)$$

$$\frac{dp_s}{dt} = e(Q + (\vec{u},\vec{E}) - \frac{1}{c}(\vec{v},\vec{G}) - \frac{1}{c}(\vec{v},[\vec{u},\vec{H}])).$$

In [5] the Lienar-Vihert potentials for case of extended space are constructed. Strength of fields $\vec{E},\vec{H},\vec{G},Q$ (appropriated to Lienar-Vihert potentials) was found. The concrete example of a particle with a mass m and charge e is considered, provided that space coordinates x, y, z of a particle are constant, but the coordinate s can vary.

The expression for electric field strengths created by a charged particle for this case was obtained

$$\vec{E} = ec\frac{c^2 - u_s^2 + R_s\dot{u}_s}{(cR_{(4)} - R_s u_s)^3}\vec{R}_{(3)}. \qquad (19)$$

Here $R_{(3)}$ is 3-dimencional distance between view points of a field and it source in a Euclidean space,
$$R_{(3)} = (x - x')^2 + (y - y')^2 + (z - z')^2$$
$R_{(4)}$ is 4-dimentional space interval which is a part of a full interval (2),
$$R_{(4)} = (x - x')^2 + (y - y')^2 + (z - z')^2 + (s - s')^2$$

and $R_s$ - distance on the 5-th coordinate S in extended space

$$R_s^2 = (s - s')^2 = c^2(\Delta t)^2 - R_{(3)}^2$$

The denominator of expression (19) $(cR_{(4)} - R_s u_s)^3$ always is more than zero. It happens, as the velocity of a charge u always is less than speed of light in a vacuum c, and the distance $R_s$ on a coordinate s always is less than a full distance $R_{(4)}$ (as a source and view point always spatially are divided).

But the numerator of expression (19) can change a sign in dependence from magnitude and sign of acceleration $\dot{u}_s$. In case, when this acceleration is rather great, i.e. the velocity of a modification of an optical denseness of a medium around a fixed charge is great, the stress of an electrical field $\vec{E}$ will change the sign. As a result particles with equal signs of charges will attract to each other, and particles with different signs of charges will push away.

As it was shown [5], that all the fields $\vec{E}, \vec{H}, \vec{G}, Q$ can change the sign in dependence with a sign and magnitude of acceleration $\dot{u}_s$. Such modification of a sign of strengths of fields (and as a result a modification of a sign of the Lorentz force) is possible to connect with a response of a radiation of these fields. This response arises, when charged particles move with acceleration.

The change of a sign at strengths in a model of extended space is of interest from that point of view that in some case under certain conditions force of interaction between particles can change the sign too. In particular, the force of an attraction of two massive particles can change to force of repulsion. It can be interpreted as manifestation of "antigravation ".

## 2. Experimental

In a model of extended space was obtained, that the electromagnetic field could be a source of a gravitational field. Besides the driven massive charged particle during deceleration can create around itself a gravitational field [1,3,4,5,7].

For experimental check of this supposition was offered and carried on described below experiment. In this experiment the probable origin of a gravitational field during deceleration of the relativistic electrons was determined on a modification of oscillations of a massive rotating pendulum.

The narrow bundle of relativistic electrons from a microthrone 1, (average power of a bundle of 450 W, energy of electrons 30 MeV) was directed on the stopping target (position 2 or 3). The target was made from tungsten. The electrons deceleration happened in a position 2 or 3.

For filing a gravitational field probably originating, during of electron deceleration near to the stopping target the special rotating pendulum placed. The pendulum was on vertical pendant 5. Pendant was made from a metal string with diameter of 1,8 mm. Pendant length was 85 sm. The pendulum could freely rotate on pendant only in a horizontal plane.

The pendulum consists of easy aluminum bar 4 (length 120 sm). On ends of a bar 4 the massive cargoes 6 and 7 of a unmagnetized material of 4 kg each were fixed, in a center the pendulum was attached by special strengthening to vertical pendant 5. The pendulum was grounding and additionally shielded from different directions by metal grid for a diminution of influence electromagnetic noise. The period of free oscillations of a pendulum was about 40 seconds.

The vertical pendant rigidity of a pendulum could vary by restriction of length of an effectively working part of the pendant. Therefore the period of oscillations could continuously be changed in limits from 40 about 27 seconds. For a diminution of influence of mechanical noise and the introductions of additional damping in oscillations of a pendulum were used two liquid dampers 10 and 11. The dampers 10 and 11 were placed near to the massive cargoes of a pendulum.

Observation of deviations of a pendulum was carried out at a special screen by the deviation of a laser ray reflected from a flat mirror 8 attached to a pendulum. For this purpose the ray from continuous He-Ne laser 9 through an optical system 12 (narrowing down an angle of a laser ray) was directed on a mirror through the special narrow channel 14 in a ferroconcrete guard 13 around of the microthrone. The ray, reflected by a mirror, registered on a screen 16 located on a distance 500 sm from a mirror by a video system 15. The video system 15 allowed remotely inspecting oscillations of a laser spot and additionally increased a sight angle in 12 times. The diameter of a focussed laser ray on a screen 16 was 0,15 mm. The maximum turn angle of a pendulum, inside which reflected ray remained in limits of the receiving channel was approximately 2 degrees. Exactitude of fixing of a turn angle by all system was $5 \cdot 10^{-4}$ degrees.

The pendulum was placed so that one of massive cargoes was near to the stopping target at a distance about 20 sm. There was also possibility to move the stopping target from one end of a pendulum (position 2) to other end (position 3). It allowed under constant parameters of all not taken into account mechanical and electromagnetic noises to change the place of stopping of an electron beam. Thus, was changed direction twisting of a pendulum under probable action of an originating gravitational radiation.

The scheme of experimental installation was placed on Fig.1.

3. Calibration measurements

For realization of measurements it was necessary to select optimum parameters of a pendulum (mass of cargoes, rigidity of pendant, magnitude of damping of oscillations). On the one hand, it is desirable, that during measurements the oscillation frequency of a pendulum was whenever possible maximum. On the other hand, the ray reflected from a mirror, should not fall outside the limit observations, limited to a diameter of the narrow observant channel in a radiation guard around the accelerator. Besides the characteristic operating time of the accelerator under a load usually was 10-15 minutes. Necessity to accumulate during this time sufficient statistical material poses restriction on a period of the oscillations and time of establishing new position of equilibrium of a pendulum in case of exterior action. All these requirements were whenever possible taken into account under the choice of final parameters of the unit.

The example of free oscillations of a pendulum in the case of availability of small mechanical vibrations from working vacuum pumps is reduced on Fig.2. – series I (the experiment was carried out 31.05.2001). On the graph are represented the amplitudes of oscillation on a screen of a laser ray (upper and lower series of values) that was reflected from a mirror attached to a pendulum, in dependence from number of oscillation and a position of an equilibrium (which was calculated from these amplitudes). The exactitude of the definition of a position of a center of a light spot was 0,1 mm. The measurements were carried out in this case involved one liquid damper and enlarged rigidity of the pendant (by restriction of effectively working pendant length). The period of free oscillations in this series was 29 seconds. Establishing of undamped oscillations of a pendulum around of the average value of equilibrium of 2,2 mm in this happened. The average amplitude of established oscillations was about 0,2 mm.

For study of the response of a pendulum on a small constant exterior force the ventilation of one of massive cargoes by very weak constant stream of an air was carried out. In this case (Fig.2, series II) the noticeable variation of oscillations of a pendulum happened already through 3-4 periods. Full establishing of the new position of equilibrium happened through 7-8 oscillations.

Both liquid dampers were involved in the other series of calibration measurements (spent 07.06.2001) when in addition pendants rigidity was reduced. The period of free oscillations of a pendulum in this case was about 40 seconds.

Adding the second damper and the diminution of pendant rigidity leads diminishing of oscillation frequency of a pendulum under action of an exterior force. On the other hand noticeable modification of a position of equilibrium of a pendulum happened already through 1-2 oscillations.

Also periodical checking of an independence of an initial average position of equilibrium of a pendulum during the time was made. For example in a series of measurements 07.06.2001 were conducted not only measurement of a position of an equilibrium prior to the beginning principal series of measurements, but also in 2 hours after a termination of principal series.

4. Experimental results and discussion.

For a period from 17.05.2001 to 07.06.2001 7 series of measurements under various operational modes of the accelerator and different parameters of a pendulum was conducted.

Before switching on and after switching off of the electronic bundle the control of position of equilibrium of a pendulum was carried out, as well as in case of calibration measurements. In this case electrical and the mechanical noises remained constant during all phase of measurements. It was achieved with the help of preliminary switching on of all devices, which were used during the measurements (water and vacuum pumps, magnetron, deflecting magnets etc.) and their cut-off only after a full termination of measurements.

At Fig.3 there are results of a measurement of an average position of a pendulum in case of position of the stopping target in a position 3 (see Fig.1). Series I and III at the graph corresponds to control measurements immediately before inclusion and in some minutes after deenergizing an electronic bundle. Series II-A and II-B reflects total oscillations of a pendulum during the moment of work of the accelerator (about 10 minutes) and some time after deenergizing a bundle. A line of a trend (average on 3 points) additionally plotted.

At Fig.4 there are results of similar experiment, which the only difference was position of the stoping target in a position 2. A line of a trend (average on 3 points) also is added.

From a qualitative comparison of lines of a trend on Fig.3 and Fig.4 it is possible to conclude, that there is correlation between inclusion of an electronic bundle and mean deviation of a pendulum from a position of equilibrium on a comparison with a control series before and after inclusion. Also was found that the direction of a deviation varies in an association with near to which of the cargoes of a pendulum the stoping target was.

The coefficients of variations that describing relative amplitude of a deviation of a pendulum for a series I, II-A, II-B, III shown on Fig.3 were also calculated. Coefficient of variation V is equal under the definition to a ratio of dispersion of some variable to magnitude of average value of this variable $V = \sigma(x)/\overline{x}$, [10]. Factor of a variation of oscillation frequency for a series II-A (moment of work of the accelerator) is equal 38 %. These exceeds coefficients of variations not only for control series I and III (18 and 23 % accordingly), but also exceeds result for a transitional series II-B: 29 % (when the electron beam was present only at the beginning of a series).

Unfortunately, for the reasons, independent from the author, at present it is difficult to improve accuracy of the experiments or to accumulate a large statistical material. So an additional evaluation by means of various methods of statistics was conducted, to prove that fact that the fixed deviations statistically are authentic.

Follow estimations was made: does the distinctions between control and principal series of measurements are statistically significant or not? The results presented below are on the base of handlings of the experimental series I, II-A, II-B, III (that were presented on Fig.3.).

The evaluation of reliability of distinctions on a Pirson criterion was made (a criterion $\chi^2$ see [10]) between a series I and III on the one hand and series II-A, B on the other hand. The theoretical reliability 99 % criterion that these series belong to various general sequences, is equal our case 13,3. Coefficient, calculated from the obtained experimental data, "$\chi^2$" has appeared equal to 30,6.

Thus by a Pirson criterion the probability of authentic statistical distinction between outcomes of control measurements of oscillations of a pendulum on a comparison with phases of an exposure of the stoping target by relativistic electrons exceeds 99 %.

In addition to parametrical significance test of distinction (criterion $\chi^2$) between control I, III (the electron beam is absent) and basic II-A, II-B (the electron beam is present) series of measurements the checking with the help of the nonparametric criterions of distinction was made.

The serial criterion of difference [10] (based on a probability evaluation of number, necessary for authentic difference of a continuous series of values of compared populations) has given 99 % reliability of probability of differences.

The criterion of Kolmogorov-Smirnov [10] (based on a comparison of series of accumulated particulars of two compared empirical populations) has given values lying between levels 95 % and 99 % of reliability of distinction of a researched experimental series.

Evaluation of magnitude of a force that can cause such a displacement of the equilibrium position of the pendulum also was made. In conducted experiments this deviation did not exceed 1-2 mm (in terms of registering scale). The calibration of the pendant gives an upper level of this force, if this force is applied on a massive cargo at the end of a pendulum - no more than $10^{-6}$ N.

## 5. Conclusion

A series of experiments on checking of a prediction about the possibility of generation of the gravitational field during deceleration of charged massive particles in substance was conducted.

As a source of charged particles the accelerator of electrons was used. The narrow bundle of relativistic electrons (average power of a bundle of 450 W, energy of electrons about 30 MeV) was directed on the stoping target from tungsten, where deceleration of the accelerated electrons happened.

The measurements have shown emerging of statistically reliable deviation of a rotating pendulum (one from which massive cargoes is near to the stoping target) in the moment of deceleration of the bundle of relativistic electrons.

Modification of direction of pendulums deflection in the case of shifting of the stopping target from one end of the pendulum to another also was fixed. Estimation of the force that calls a deviation of a pendulum has an upper level of $10^{-6}$ N.

Certainly, these first experimental results on checking of predictions which were made on the base of a model of extended space, have a preliminary character and require much more substantial check, as will be further basis for the future experiments.

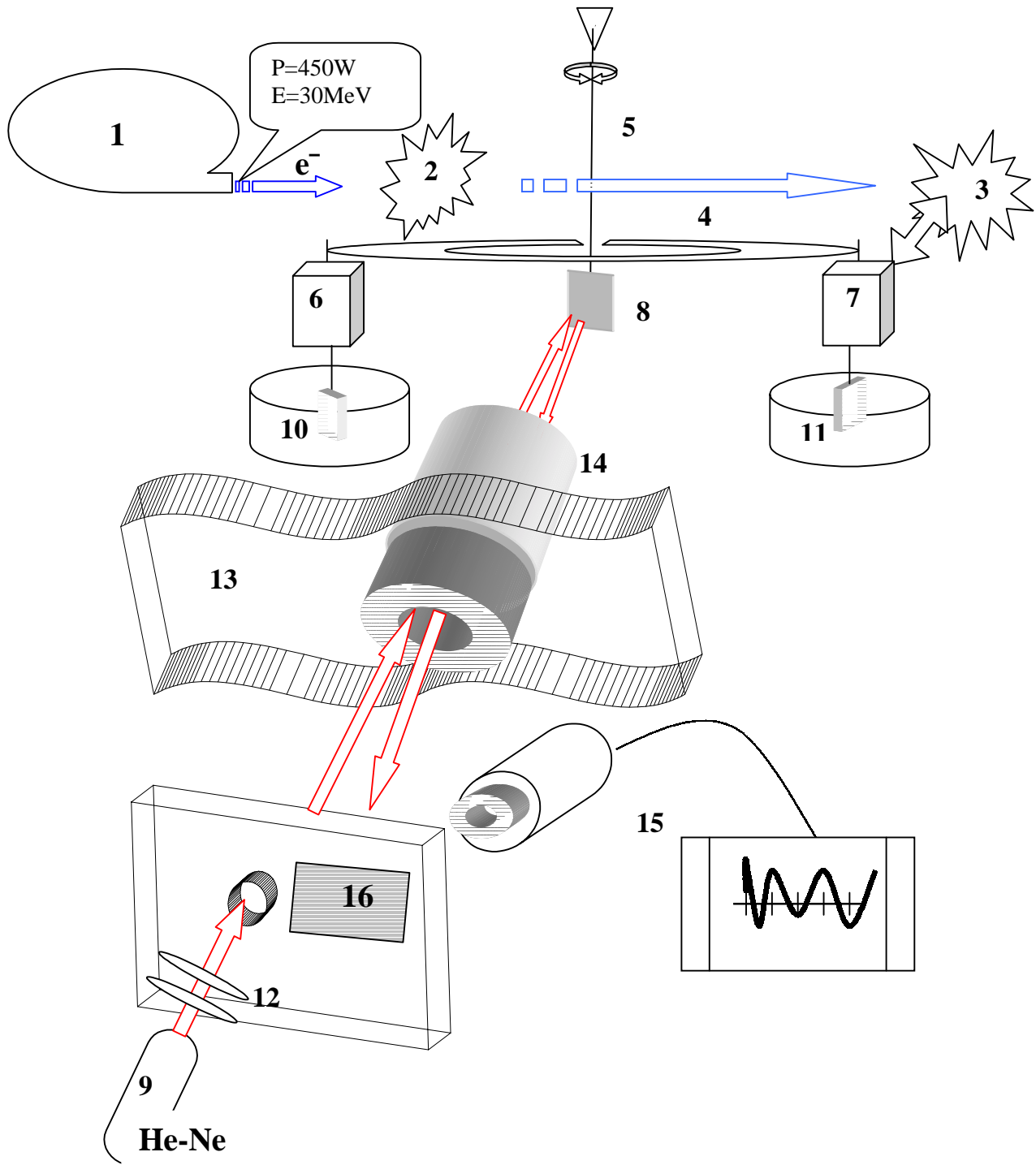

Fig.1

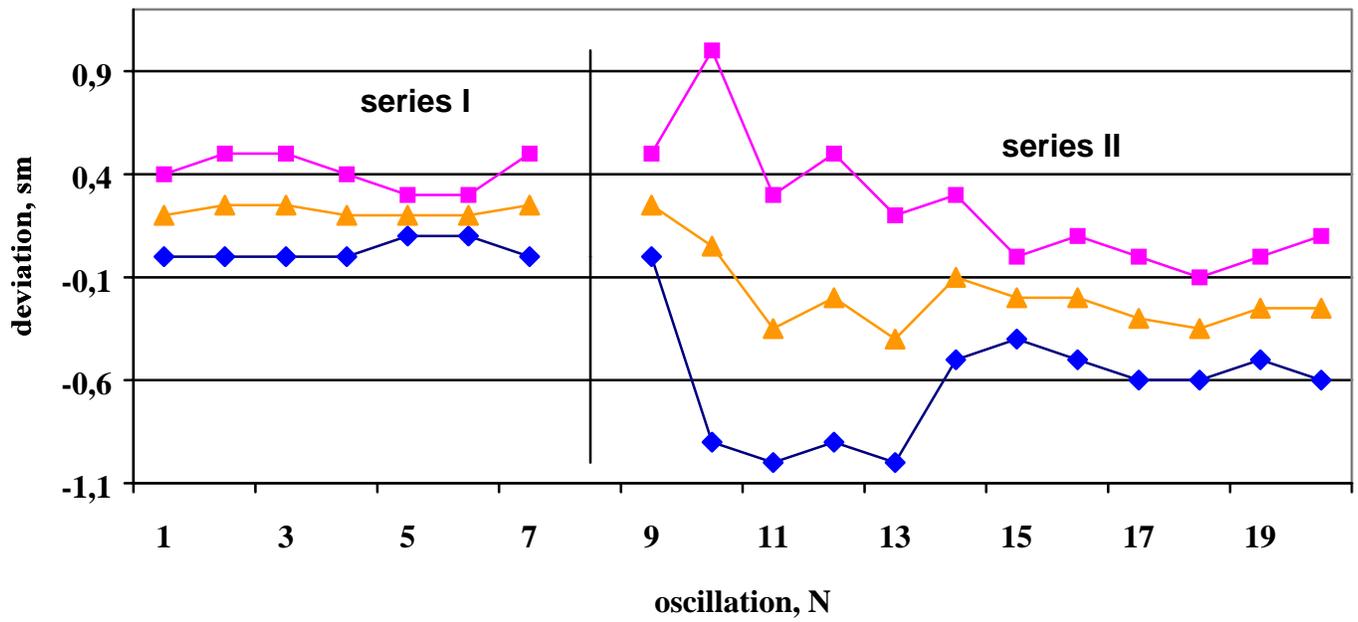

**Fig.2. Calibration measurements 31.05.2001**

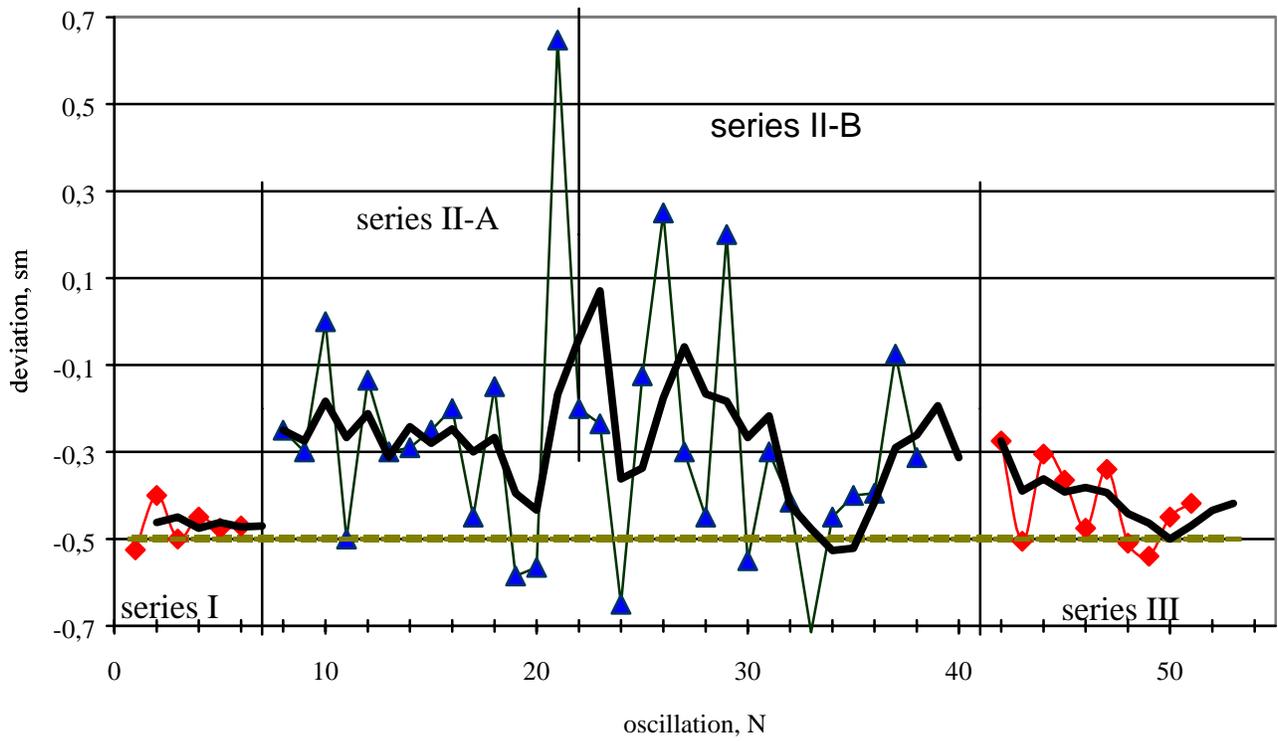

**Fig.3. Stopping target in position 3**

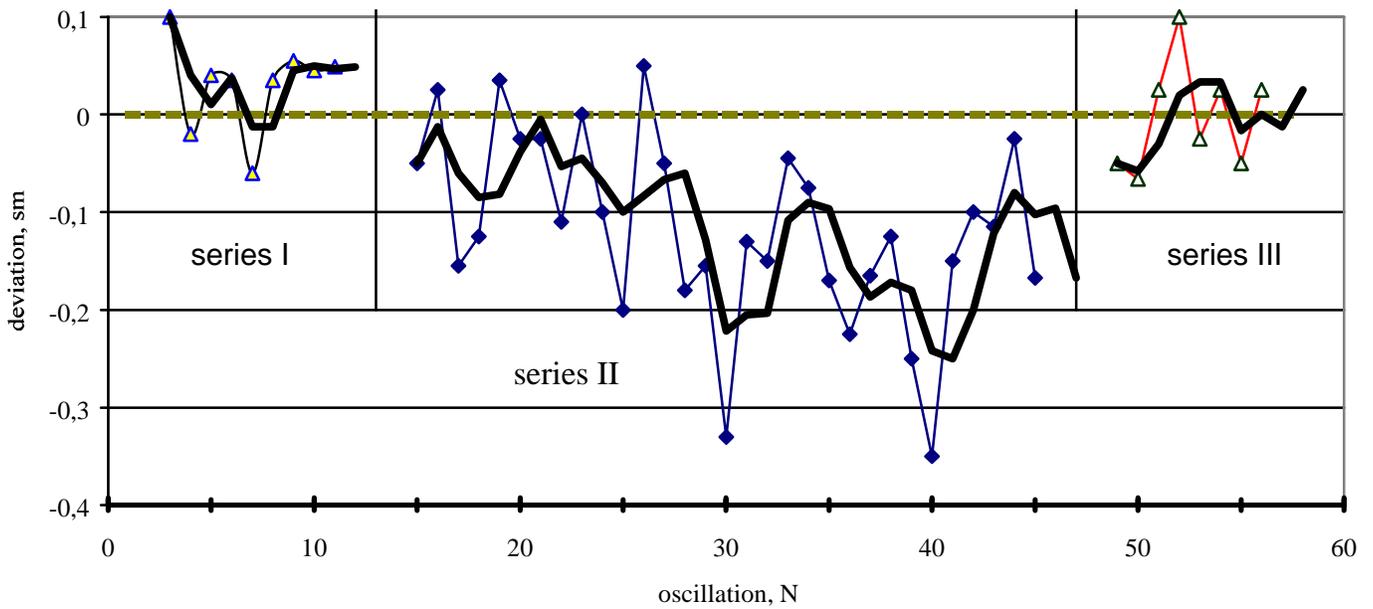

**Fig.4. Stopping target in position 2**